\documentclass[aps,twocolumn,floatfix,showpacs]{revtex4}
\usepackage{graphicx}
\usepackage{amsmath}
\usepackage{amssymb}
\usepackage{bm}

\begin{document}
\title{Self-organization in multimode microwave phonon laser (phaser):\\
experimental observation of spin-phonon cooperative motions}
\author{D.~N.~Makovetskii}%
\email{makov@ire.kharkov.ua} \affiliation{Institute for Radio-Physics
and Electronics of National Academy of Sciences of Ukraine}

\date{March 6, 2003}

\begin{abstract}
An unusual nonlinear resonance was experimentally observed in a ruby
phonon laser (phaser) operating at 9 GHz with an electromagnetic
pumping at 23 GHz. The resonance is manifested by very slow cooperative
self-detunings in the microwave spectra of stimulated phonon emission
when pumping is modulated at a superlow frequency (less than 10 Hz).
During the self-detuning cycle new and new narrow phonon modes are
sequentially ``fired'' on one side of the spectrum and approximately
the same number of modes are ``extinguished'' on the other side, up to
a complete generation breakdown in a certain final portion of the
frequency axis. This is usually followed by a short-time refractority,
after which the generation is fired again in the opposite (starting)
portion of the frequency axis. The entire process of such cooperative
spectral motions is repeated with high degree of regularity. The
self-detuning period strongly depends on difference between the
modulation frequency and the resonance frequency. This period is
incommensurable with period of modulation. It increases to very large
values (more than 100 s) when pointed difference is less than 0.05 Hz.
The revealed phenomenon is a kind of global spin-phonon
self-organization. All microwave modes of phonon laser oscillate with
the same period, but with different, strongly determined phase shifts
--- as in optical lasers with antiphase motions.
\end{abstract}


\maketitle

Phonon amplification by stimulated emission of radiation was predicted
theoretically in \cite{TheorEarly} and experimentally observed in
microwave range by several groups \cite{AmplifEarly} in 1960--1970-th.
This phenomenon is very similar to the usual paramagnetic maser gain of
electromagnetic field \cite{Siegman-1964,TP-1991}. But if the phonon
gain is large enough to exceed the phonon losses in the solid-state
resonator, the self-excitation of \textit{laser-like} phonon emission
is possible \cite{GenerTucker,SSC-1974,JETP-1977}. The wavelength of
generated microwave-frequency phonons in such phonon laser lies usually
in optical or near-infrared range (due to very small velocity of sound
in crystals, which is about 5 orders less than the light velocity). In
this sence microwave phonon laser (phaser) is more close relative of
the optical laser, than terahertz phonon laser (saser), having much
shorter wavelength (see \cite{Terahertz} and references therein).

Almost all early experiments with microwave phonon lasers
\cite{GenerTucker,SSC-1974,JETP-1977} were carried out in autonomous
regime, when the control parameters of the active system remain
unchanged during the whole time of measurement. Various regular and
chaotic processes of generation of microwave phonons in a multimode
\textit{nonautonomous} phaser was experimentally observed and studied
\cite{Chaotic} for $\omega_m \approx \omega_R \approx$ 20--300~Hz,
where $\omega_m$ is the pumping modulation frequency, $\omega_R$ is the
relaxation frequency of a nonequilibrium autonomous acoustic system
($\omega_R$ depends on pumping level). The existence of a single
dominating component $\exp(i \omega_R t)$ for the transient processes
in the autonomous phonon laser reflects the collective character of a
multimode stimulated emission \cite{LaminarTheory} in the quantum
generators of class $B$, an acoustic analog of which is represented by
a system studied in \cite{Chaotic}. As a result, a nonlinear dynamics
of the integral intensity of the multimode stimulated phonon emission
$J_{\Sigma}(t)$ in a nonautonomous phonon laser operating at $\omega_m
\approx \omega_R$ \cite{Chaotic} is satisfactorily described by a
single-mode (or even point-like) model.

Outside the region of the nonlinear low-frequency resonance indicated
above, the pronounced dependence of the mode variables on the scalar
order parameter $J_{\Sigma}$ is violated. Even in a two-mode regime of
the class $B$ quantum generator, evolution of the partial components
$J_{1,2}(t)$ qualitatively differs from $J_{\Sigma}(t)$
\cite{Georgiou-1994}. In particular, the system may exhibit
manifestations of a new nonlinear resonance at a frequency $\omega_L
\ll \omega_R$, which is related to the phonon intermode energy exchange
through spin-system.

In this study, a nonlinear superlow-frequency resonance in a ruby
phaser was experimentally observed for the first time at $\omega_m
\approx \omega_L \lesssim 10$~Hz. This resonance leads to a manifold
narrowing of the microwave spectrum of stimulated phonon emission and
to the appearance of very slow, highly organized self-detunings of
phaser generation at still lower frequencies depending on the parameter
$\Delta_L \equiv \omega_m - \omega_L$. Some preliminary results on
superlow-frequency resonance were reported in \cite{LaminarEarly}, and
short version of this study was published in \cite{TPL-2001} and
repoted in \cite{MSMW-2001}.

The main element of the phonon laser studied is a solid-state microwave
acoustic Fabry-P\'erot resonator (AFPR) made of a pink ruby crystal
${\mathrm{Cr^{3+} : Al_{2}O_{3}}}$ (with impurities of
${\mathrm{Cr_{2x}^{3+}}}$ in ${\mathrm{Al_{2(1 - x)}O_3}}$, where
${\mathrm{x}} \approx 3 \times 10^{-4}$). This AFPR was made of ruby
single crystal grown by the Verneuil method and cut to form of a round
rod of 2.6~mm diameter and 17.6~mm length. Measured acoustic $Q$-factor
of the AFPR is $Q_N \approx 10^6$ at 9--10~GHz.

One of the resonator's acoustic mirrors was coated with a textured
piezoelectric ${\mathrm{ZnO}}$ thin film capable of detecting
longitudinal acoustic ($LA$) oscillations in AFPR. These oscillations
were excited in the ruby crystal lattice due to the stimulated phonon
emission at spin transitions $E_3 \rightarrow E_2$ of
${\mathrm{Cr^{3+}}}$ paramagnetic ions. Inversion of populations of
spin-levels $E_3,\; E_2$ was provided by simultaneous (push-pull)
pumping of the transitions $E_1 \rightarrow E_3$ and $E_2 \rightarrow
E_4$. Here $E_i \; (i = 1..4)$ are the ${\mathrm{Cr^{3+}}}$ ground spin
multiplet levels.

\begin{figure}
  \centering
  \includegraphics{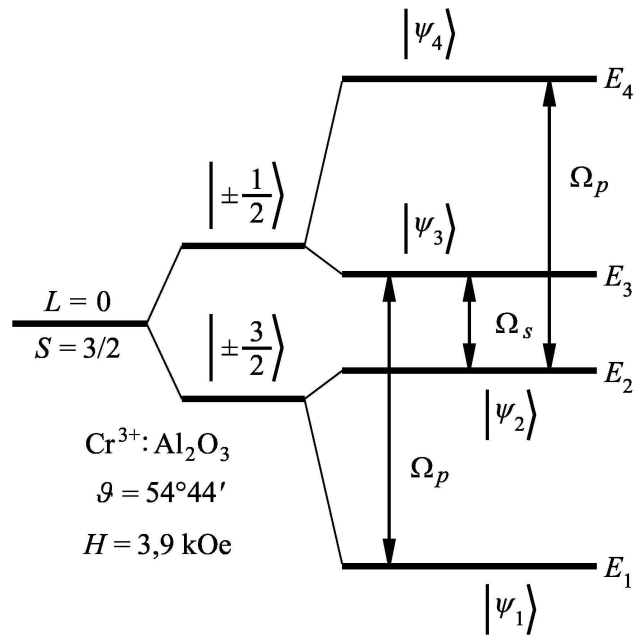}
  \caption{\label{fg:1}Push-pull scheme of the energy levels
  of $\mathrm{Cr^{3+}}$ active centers in pink ruby. The lowest spin quadruplet
  (orbital quantum number $L=0$, spin quantum number $S=3/2$) is splitted
  by electric crystal field into two doublets $\left | \pm\frac{1}{2} \right
  \rangle$ and $\left |  \pm\frac{3}{2} \right \rangle$.
  Static magnetic field $\bm{H}$ splits these doublets giving the symmetric
  scheme of energy levels at $\vartheta = 54^{\circ}44^{\prime}$, were
  $\vartheta$ is the angle between $\bm{H}$ and ruby optical axis
  $\bm{\mathrm{C}}_3$. Energy levels $E_i$ and wave functions
  $\left | \psi_i \right \rangle$ are the eigenvalues and eigenfunctions
  of the ${\mathrm{Cr^{3+} : Al_{2}O_{3}}}$ spin-hamiltonian (see e.g.
  \cite{Siegman-1964}). At $H \approx 3.9$~kOe the push-pull pumping frequency
  $\Omega_P \equiv (E_4 - E_2)/\hbar \equiv (E_3 - E_1)/\hbar \approx 23$~GHz
  and the signal frequency $\Omega_S \equiv (E_3 - E_2)/\hbar \approx 10$~GHz.}
\end{figure}

The experiments were performed at $\omega_m =$ 2--20~Hz with the ruby
phaser having push-pull electromagnetic pumping at a frequency of
$\Omega_P = 2.3 \times 10^{10}$~Hz (see FIG.~\ref{fg:1}). The mode
frequencies $\Omega_N$ of the generated $LA$-phonons occur near a
maximum of the inverted acoustic paramagnetic resonance (APR) line
$\Omega_S = 9.12 \times 10^9$~Hz. Since the longitudinal sound velocity
$V_{LA}$ in the system studied was $V_{LA} = V_{LA}(\bm{k} \parallel
\bm{\mathrm{C}}_3) \approx 10^6$~cm/s, the wavelengths $\lambda_N$ of
the generated microwave acoustic modes were close to the
optical-wavelength ones: $\lambda_N \approx 1 \; \mu$m
($\bm{\mathrm{C}}_3$ is the trigonal axis of ruby crystal lattice). The
intermode spacing was $\Omega_N - \Omega_{N-1} = 3.1 \times 10^5$~Hz
and the total number of modes reached $N_0^{\text{max}}= 23$ at a
nonmodulated pumping power of $P = 1.2 \times 10^{-2}$~W and the
quality factor of electromagnetic pumping cavity $Q_P = 10^4$.

\begin{figure}
  \centering
  \includegraphics{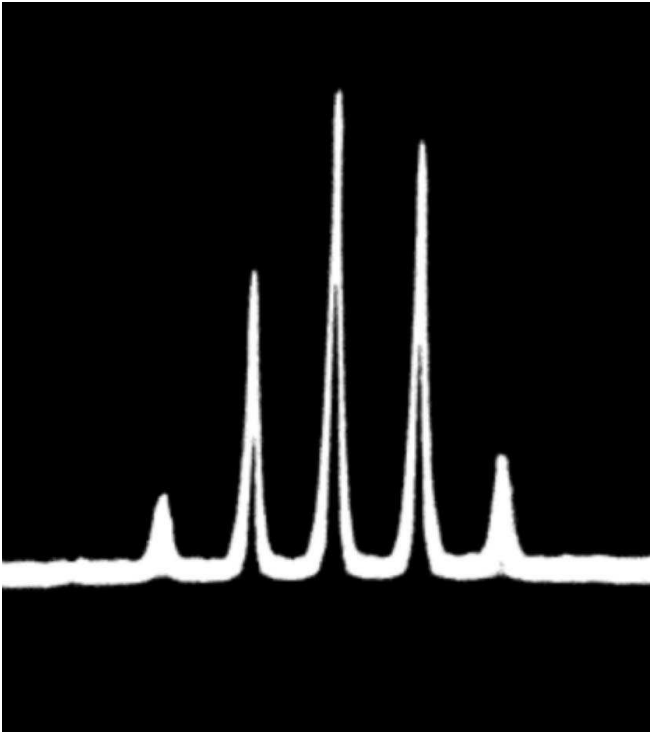}
  \caption{\label{fg:2}Power spectrum of phonon induced emission in
  autonomous microwave phonon laser (phaser) at $\theta = 1.7$~K.
  Exceeding of the pump threshold is small (about 1~dB).
  The intermode distance is 310~KHz, which is equal to
  the splitting of the AFPR eigenmodes for the $LA$ oscillations
  with wave vector $\bm{k} \parallel \bm{\mathrm{C}}_3$.}
\end{figure}

All the experiments were performed by measuring the microwave spectra
of the $LA$-oscillations at temperatures of $\theta = 1.7 - 4.2$~K in
static magnetic fields $H$ within the interval $H = [H^{(-)},
H^{(+)}]$, where $ H^{(\pm)} = H_0 \pm |{\Delta}H|_{\text{max}}$, $H_0
= 3920$~Oe, ${\Delta}H = H - H_0$, $|{\Delta}H|_{\text{max}} = 50$~Oe.
The magnetic field vector $\bm{H}$ was oriented at an angle of
$\vartheta = 54^{\circ}44^{\prime}$ with respect to the axis
$\bm{\mathrm{C}}_3$ of the ruby crystal, which was necessary to ensure
the condition of the push-pull pumping $E_3 - E_1 = E_4 - E_2$
(FIG.~\ref{fg:1}). A typical power spectrum of microwave phonon
emission in \textit{autonomous} ruby phonon laser is shown at
FIG.~\ref{fg:2}.

A special feature distinguishing this experiment from those described
in \cite{Chaotic} was the use of a pulse lock-in method for detecting
the microwave acoustic spectra. The essence of the method is as
follows. A stimulated phonon emission generated in the crystal is
converted by the piezoelectric film into an electromagnetic signal
transmitted to a microwave spectrum analyzer. In the normal state, a
beam of the input oscilloscope in the spectrum analyzer is shut off.
The oscilloscope is periodically open by strobe pulses only within very
short time intervals ${\Delta}t_{\text{str}} \ll T_m \equiv
{2\pi}/{\omega_m}$, with a strobing period $T_{str}$ being equal
exactly to the period of external pumping modulation $T_m$. At each
lock-in time instant, the beam displayed a set of points on the
oscilloscope screen indicating an instantaneous frequency distribution
of the phonon mode intensity.

The shutter of a camera making photographs of the oscilloscope screen
is open within a certain registration time interval $t_{\text{reg}}$,
which is set as $t_{\text{reg}} = nT_m$ ($n \gg 1$ being an integer).
Therefore, each camera shot represents a series of superimposed point
sets. If the period of intensity oscillations for all of the generated
phonon modes coincides with the external modulation period $T_m$, a
spectrum measured in this stroboscopic regime must contain a singe
point representing each mode.

A different situation takes place when the intensity oscillations of
the generated phonon modes are not in phase with the modulation factor.
For example, if the oscillation period is doubled simultaneously for
all the modes, the number of points on the image would double as well.
An especially illustrative pattern observed for the stimulated emission
periodicity breakage, whereby the distribution of points on the image
becomes chaotic, the intermode correlation is lost, etc.

It is important to note that both qualitative and quantitative
information can be obtained concerning the degree of spectral ordering,
the character of regular and irregular detunings in the mode structure,
etc. The pro-posed method is essentially a generalization of the
well-known Poincar\'e cross-sections method (see, e.g.,
\cite{Schuster-1984}) to the case of a multimode system. A highly
illustrative character --- one of the main advantages of the Poincar\'e
method --- is fully inherited in the proposed approach.

\begin{figure}
  \centering
  \includegraphics{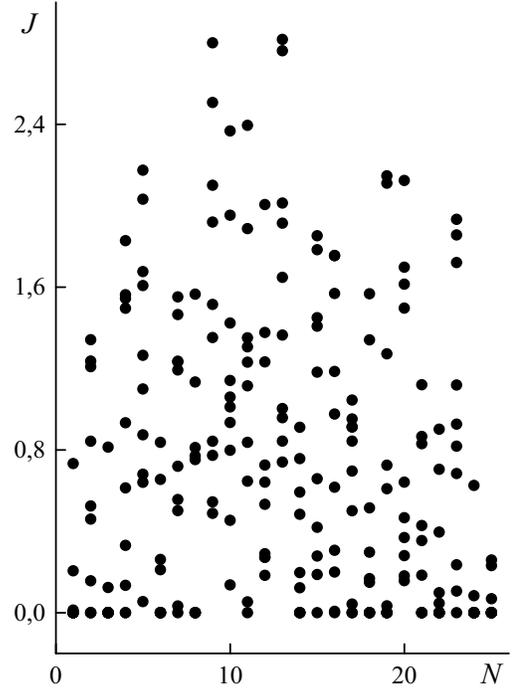}
  \caption{\label{fg:3}Stroboscopic power spectrum of stimulated microwave
  phonon emission in nonautonomous ruby phaser with the electromagnetic
  pumping modulated at the maximum of low-frequency nonlinear resonance:
  $\omega_m=9.79$~Hz. Here $N$ is the microwave phonon mode number;
  $J$ is the instantaneous intensity of the phonon stimulated emission
  normalized to the intensity of the central (maximum power) mode of the
  autonomous generation. $H=3.92$~kOe; $\theta = 1.8$~K.}
\end{figure}

\begin{figure}
  \centering
  \includegraphics{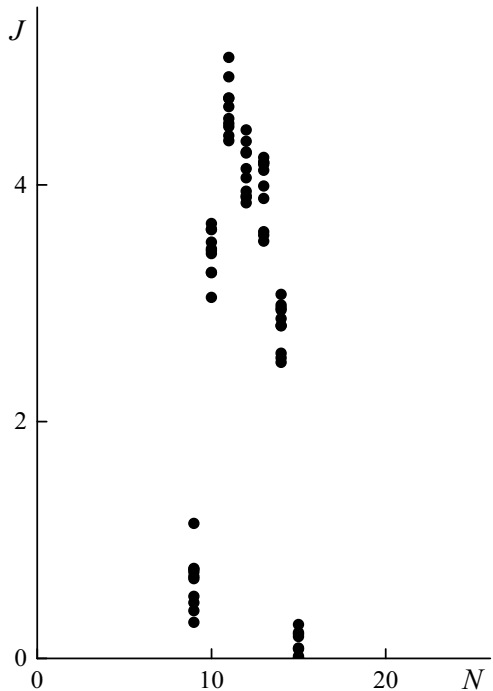}
  \caption{\label{fg:4}The same as at previous figure, but with the
  electromagnetic pumping modulated in the region of superlow-frequency
  nonlinear resonance: $\omega_m=9.79$~Hz.}
\end{figure}

Figures \ref{fg:3} and \ref{fg:4} show typical stroboscopic spectra of
the stimulated phonon emission measured at $\omega_m = 94.0$~Hz (i.e.,
in the region of a low-frequency resonance $\omega_m \approx \omega_R$)
and at $\omega_m = 9.79$~Hz (at the maximum of the superlow-frequency
resonance observed for the first time). The measurements were performed
for $n = 10$. As is clearly seen, a low-frequency destabilization of
the generation process is manifested primarily by a strong chaotic
amplitude modulation of the phonon modes (FIG.~\ref{fg:3}). This is
accompanied by certain broadening of the spectrum ($N_R^{(\text{max})}
= 25$) as compared to that observed in the autonomous regime. The
variation of $\omega_m$ from 70 to 200 Hz does not lead to qualitative
changes of the general pattern (in agreement with the data reported
previously for $J_{\Sigma}(t)$ \cite{Chaotic}).

On the contrary, a superlow-frequency destabilization of the generation
process is manifested by a pronounced (manifold) narrowing of the
spectrum ($N_L^{(\text{max})} = 7$) on the background of a
significantly lower mode amplitude automodulation (FIG.~\ref{fg:4}).
However, a much more pronounced feature is the appearance of
self-detunings in the microwave phonon spectrum for $\omega_m$
deviating from the maximum of the superlow-frequency resonance at
$\omega_m = 9.79$~Hz. The self-detuning process is manifested by motion
of the region of localization of this relatively narrow mode cluster
along the frequency axis at a retained localization (width) of each
mode. In other words, new modes are sequentially ``fired'' on one side
of the cluster and approximately the same number of modes are
``extinguished'' on the other side, up to a complete generation
breakdown in a certain final portion of the frequency axis. This is
usually followed by a short-time refractority, after which the
generation is fired again in the opposite (starting) portion of the
frequency axis and the entire process of collective motions is multiply
repeated at a period of $T_{SD} \gg T_m$.

\begin{figure}[t]
  \centering
  \includegraphics{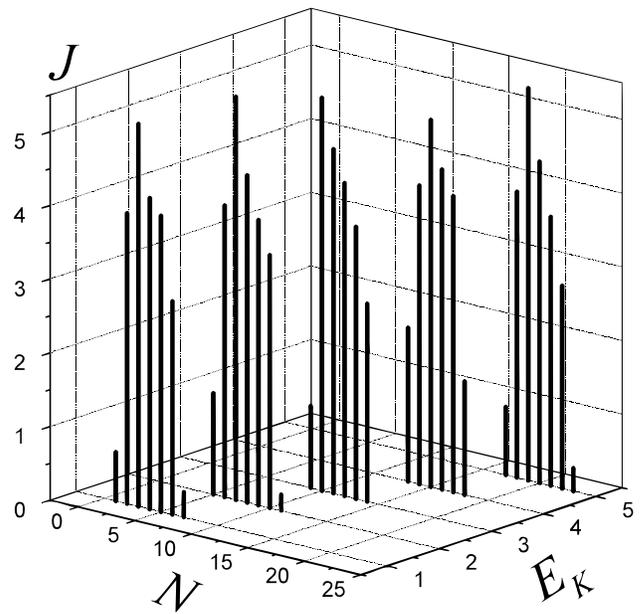}
  \caption{\label{fg:5}A series of stimulated phonon emission spectra
  measured for small detunings ($\Delta_L = -0.23$~Hz) of the modulation
  frequency relative to the superlow-frequency resonance maximum.
  $E_K$ is the number of sequential exposure for the spectra taken at
  an interval of $T_E \approx 2.5$~s (other notations as in
  previous figures.}
\end{figure}

FIG.~\ref{fg:5} shows a series of stimulated phonon emission spectra
measured for $\Delta_L = -0.23$~Hz and $\Delta_H = 0$. A time interval
of $T_E \approx 2.5$~s between sequential exposures $E_K$ in
FIG.~\ref{fg:5} is selected to be much greater than $T_m$. For the
$\Delta_L$ value indicated above, the self-detuning period is $T_{SD}
\approx 14$~s. The results of experiments showed that the $T_{SD}$
value strongly depends on both absolute value and sign of $\Delta_L$
and varies from a few tenths of a second for $|\Delta_L| \approx 1$~Hz
up to very large values $T_{SD} > 100$~s for $|\Delta_L| < 0.05$~Hz.
The direction of motion of the mode cluster is determined by the sign
of $|\Delta_L|$: for the virtual cluster maximum position denoted by
$\Omega_B$, the direction is such that $\text{sgn} (d \Omega_B / dt) =
- \text{sgn}{\Delta_L}$.

An analogous character of self-detunings in the microwave phonon
spectrum in the region of the super-low-frequency resonance is retained
on the whole in a rather broad range of $\Delta_H$ . Moreover, in the
entire region of $|\Delta_H| < 10$~Oe, even the $\omega_L$ value
remains virtually the same (close to 9.8~Hz). Only for a still greater
detuning of the spin system with respect to the magnetic field (under
these conditions, even an autonomous phonon laser may exhibit some
insufficiently studied generation features \cite{TPL-1994}), the
resonance frequency $\omega_L$ begins to drop significantly
(approximately by half for $|\Delta_H| < 60$~Oe). Here the stimulated
phonon emission intensity drops by more than one order of magnitude
because of a decrease in the level of pumping (which passes now via a
wing of the paramagnetic resonance line). A change in the sign of
$|\Delta_H|$ does not modify the self-detuning process (in contrast to
the case of changing the sign of $\Delta_L$ considered above).

On the whole, the dynamics of the collective motions described above is
somewhat analogous to that of autowaves \cite{Mikhailov-1990}. Indeed,
the observed moving spectral structures (like the usual autowaves) are
independent (within certain limits) of perturbations in the control
parameters and probably result from a self-organization in the open
dissipative system of a phonon laser. The energy of external excitation
(in our case, a pumping field) is spent for maintaining this state,
while selective external factors only switch on certain internal
processes changing the system behavior. Although the motions observed
in our experiments occur in the spectral ``space'', these phenomena
naturally reflect the corresponding processes in the real physical
space of a distributed active system of the phonon laser studied.

It should be emphasized that the above-described self-detunings in the
phonon spectra of a nonautonomous phaser, as well as the recently
observed nonlinear phenomena in electromagnetic masers \cite{TP-1999}
and autonomous acoustic quantum generators \cite{TPL-1994, RE-1999},
possess an essentially nonthermal nature. Although the lower
characteristic frequencies of the motions observed in this study fall
within a millihertz frequency range, all these motions are related to
the microwave field self-action through a spin system of a paramagnet
in the absence of the so-called phonon bottleneck effect
\cite{Antipin-1978} (and of the resulting phonon avalanche
\cite{Golenishchev}). Therefore, no overheating of the stochastic
phonon subsystem takes place in the spectral range of $10^8$~Hz
corresponding to the APR linewidth on the $E_2 \leftrightarrow E_3$
transition in pink ruby crystal at low temperatures. The observed
frequencies of the chaotic amplitude modulation occur below $10^3$~Hz,
while the microwave modes of the stimulated phonon emission remain very
broad ($\Delta\Omega_N < 10^{-7} \Omega_N$) for both $\omega_m \approx
\omega_R$ and $\omega_m \approx \omega_L$. In this respect, behavior of
the studied system qualitatively differs from that of the optical
quantum generators of class $A$ featuring a large-scale breaking of
coherency (by the hydrodynamical turbulence type \cite{Weiss-1999}) of
the stimulated emission.

In conclusion, the revealed phenomenon is a kind of global spin-phonon
self-organization, which is the result of superlow-frequency resonant
destabilization of stationary phaser generation. All microwave modes of
phonon laser oscillate with the same period, but with different,
strongly determined phase shifts --- as in optical class $B$ lasers
with antiphase motions (see \cite{LaminarTheory} and references
therein). Several features of observed cooperative alternation of
resonant $LA$ modes in phonon laser may be described by simple balance
model of active medium with spatial hole burning \cite{LaminarTheory},
but detailed theory of the self-organization in phaser must include
into consideration the electron-nuclear magnetic interactions
\cite{Abragam-1982}. These interactions dramatically modify the
processes of near-resonant energy transfer in ruby phaser
\cite{JETP-1977,FTT-1982} due to nuclear spin polarization under
conditions of electron spin-system saturation.

{\textbf{Acknowledgments}}. The author is grateful to E.~D.~Makovetsky
(Physical Optics Chair, Kharkov V.N. Karazin National University) and
S.~D.~Makovetskiy (Department of Computer Sciences, Kharkov National
University of Radioelectronics) for their valuable help with computer
processing of the experimental data.


\begin{thebibliography}{99}

\bibitem{TheorEarly}
%
%\bibitem{Townes-1960}
 C.~H.~Townes,
 {\textit{Quantum Electronics}} (Columbia Univ. Press, New York,
 1960), pp.~405-409;
%
%\bibitem{Kopvillem-1961}
 U.~Kh.~Kopvillem and V.~D.~Korepanov,
 Zh. Experim. Teor. Fiz. {\textbf{41}}(1), 211 (1961), {\textit{in Russian}};
%
%\bibitem{Kittel-1961}
 C.~Kittel,
 Phys. Rev. Lett. {\textbf{6}}, 449 (1961).

\bibitem{AmplifEarly}
%
%\bibitem{Tucker-1961}
 E.~B.~Tucker,
 Phys. Rev. Lett. {\textbf{6}}, 547 (1961);
%
%\bibitem{Shiren-1965}
 N.~S.~Shiren,
 Appl. Phys. Lett. {\textbf{7}}, 142 (1965);
%
%\bibitem{Peterson-1969}
 P.~D.~Peterson and E.~H.~Jacobsen,
 Science {\textbf{164}}, 1065 (1969);
%
%\bibitem{Doklady-1974}
 E.~M.~Ganapolskii and D.~N.~Makovetskii,
 Sov. Phys. -- Doklady {\textbf{19}}, 433 (1975)
 [translation of: Doklady AN SSSR {\textbf{217}}(2),
 303 (1974)].

\bibitem{Siegman-1964}
 A.~E.~Siegman,
 {\textit{Microwave solid-state masers}} (McGraw Hill,
 New York etc., 1964).

\bibitem{TP-1991}
 D.~N.~Makovetskii and K.~V.~Vorsul,
 Sov. Phys. -- Tech. Phys. {\textbf{36}}, 50 (1991)
 [translation of: Zh. Tekh. Fiz. {\textbf{61}}(1), 86
 (1991)].

\bibitem{GenerTucker}
%
%\bibitem{Tucker-1964}
 E.~B.~Tucker,
 {\textit{Quantum Electronics: Proc. 3-rd Int. Congress,
 Vol.~2}}, edited by P.~Grivet and N.~Bloembergen (Dunod Editeur,
 Paris; Columbia Univ. Press, New York, 1964), pp.~1787--1800;
%
%\bibitem{Tucker-1966}
 E.~B.~Tucker,
 {\textit{Physical Acoustics, Vol.~4A}}, edited
 by W.~P.~Mason (Academic Press, New York and London, 1966), p.~47;

\bibitem{SSC-1974}
 E.~M.~Ganapolskii and D.~N.~Makovetskii,
 Solid State Commun. {\textbf{15}}, 1249 (1974);

\bibitem{JETP-1977}
 E.~M.~Ganapolskii and D.~N.~Makovetskii,
 Sov. Phys. -- JETP {\textbf{45}}, 106 (1977)
 [translation of: Zh. Experim Teor. Fiz. {\textbf{72}}(1), 203
 (1977)].

\bibitem{Terahertz}
%
%\bibitem{Camps-2000}
 I.~Camps and S.~S.~Makler,
 Solid State Commun. {\textbf{116}}, 191 (2000),
 \eprint{cond-mat/0101041};
%
%\bibitem{Camps-2001}
 I.~Camps, S.~S.~Makler, H.~M.~Pastawski, and L.~E.~F.~Foa~Torres,
 \eprint{cond-mat/0101043};
%
%\bibitem{Sanders-2001}
 G.~D.~Sanders, C.~L.~Stanton, and C.~S.~Kim,
 \eprint{cond-mat/0101459};
%
%\bibitem{Lozovik-2001}
 Yu.~E.~Lozovik, S.~P.~Merkulova, and I.~V.~Ovchinnikov,
 \newblock Phys. Lett. A {\textbf{282}}, 407 (2001).

\bibitem{Chaotic}
%
%\bibitem{RE-2001}
 D.~N.~Makovetskii,
 Radiofizika i Elektronika (Proc. of Instute for Radiophysics and
 Electronics of Nat. Acad. Sci. of Ukraine) {\textbf{6}}(1), 124
 (2001), {\textit{in Russian}};
%
%\bibitem{TP-1989}
 E.~M.~Ganapolskii and D.~N.~Makovetskii,
 Sov. Phys. -- Tech. Phys. {\textbf{34}}, 1220 (1989)
 [translation of: Zh. Tekh. Fiz. {\textbf{59}}(10), 202
 (1989)];
%
%\bibitem{TP-1992}
 E.~M.~Ganapolskii and D.~N.~Makovetskii,
 Sov. Phys. -- Tech. Phys. {\textbf{37}}, 218 (1992)
 [translation of: Zh. Tekh. Fiz. {\textbf{62}}(2), 187
 (1992)].

\bibitem{LaminarTheory}
%\bibitem{Wiesenfeld-1992}
 K.~Wiesenfeld, C.~Bracikowski, G.~James, and R.~Roy,
 Phys. Rev. A {\textbf{65}}, 1749 (1990);
%
%\bibitem{Nguyen-1997}
 B.~A.~Nguyen and P.~Mandel,
 Opt. Commun. {\textbf{138}}, 81 (1997);
%
%\bibitem{Vladimirov-1999}
 A.~G.~Vladimirov, E.~A.~Viktorov, and P.~Mandel,
 Phys. Rev. E {\textbf{60}}, 1616 (1999).

\bibitem{Georgiou-1994}
 M.~Georgiou, P.~Mandel, and K.~Otsuka,
 IEEE J. Quantum Electronics, {\textbf{30}}, 854 (1994).

\bibitem{LaminarEarly}
%
%\bibitem{SSC-1994}
 E.~M.~Ganapolskii and D.~N.~Makovetskii,
 Solid State Commun. {\textbf{90}}, 501 (1994);
%
%\bibitem{UFZh-1993}
 E.~M.~Ganapolskii and D.~N.~Makovetskii,
 Ukrainian Phys. Journ. {\textbf{38}}(2), 222 (1993) {\textit{in
 Russian}}.

\bibitem{TPL-2001}
 D.~N.~Makovetskii,
 Tech. Phys. Lett. {\textbf{27}}(6), 511 (2001)
 [translation of: Pis'ma Zh. Tekh. Fiz. {\textbf{27}}(12), 57
 (2001)].

\bibitem{MSMW-2001}
 D.~N.~Makovetskii,
 {\textit{Proc. 4-th Int. Symp.: Physics and Engineering
 of Millimeter and Submillimeter Waves --- MSMW'2001 (June 4--9, 2001),
 Vol.~2}} (Kharkov, 2001), pp.762--764.

\bibitem{Schuster-1984}
 H.~G.~Schuster,
 {\textit{Deterministic chaos. An introduction}} (Physik-Verlag,
 Veinheim, 1984).

\bibitem{TPL-1994}
 E.~M.~Ganapolskii and D.~N.~Makovetskii,
 Tech. Phys. Lett. {\textbf{20}}(11), 854 (1994)
 [translation of: Pis'ma Zh. Tekh. Fiz. {\textbf{20}}(21),
 65 (1994)].

\bibitem{Mikhailov-1990}
 A.~S.~Mikhailov,
 Foundations of Synergetics I. Distributed Active
 Systems (Springer, Berlin, 1994).

\bibitem{TP-1999}
 D.~N.~Makovetskii, A.~A.~Lavrinovich, and N.~T.~Cherpack,
 Tech. Phys. {\textbf{44}}, 570 (1999)
 [translation of: Zh. Tekh. Fiz. {\textbf{69}}(5), 101
 (1999)].

\bibitem{RE-1999}
 D.~N.~Makovetskii,
 Radiofizika i Elektronika (Proc. of Instute for Radiophysics and
 Electronics of Nat. Acad. Sci. of Ukraine) {\textbf{4}}(2), 91
 (1999), {\textit{in Russian}}.

\bibitem{Antipin-1978}
 A.~A.~Antipin, L.~D.~Livanova, and A.~A.~Fedii,
 Sov. Phys. -- Solid State {\textbf{20}}, 1030 (1978)
 [translation of: Fiz. Tverd. Tela (Leningrad) {\textbf{20}},
 1783 (1978)].

\bibitem{Golenishchev}
%
%\bibitem{Golenishchev-1977}
 V.~A.~Golenishchev-Kutuzov, V.~V.~Samartsev, N.~K.~Solovarov,
 and B.~M.~Khabibullin,
 {\textit{Magnetic Quantum Acoustics}} (Nauka,
 Moscow, 1977), {\textit{in Russian}};
%
%\bibitem{Golenishchev-1988}
 V.~A.~Golenishchev-Kutuzov, V.~V.~Samartsev, and B.~M.~Khabibullin,
 {\textit{Pulsed Optical and Acoustical Spectroscopy}}
 (Nauka, Moscow, 1988), {\textit{in Russian}}.

\bibitem{Weiss-1999}
 C.~O.~Weiss, M.~Vaupel, K.~Staliunas, G.~Slekys, and V.~B.~Taranenko
 Appl. Phys. B {\textbf{68}}, 151 (1999).

\bibitem{Abragam-1982}
 A.~Abragam and M.~Goldman,
 {\textit{Nuclear magnetism. Order and Disorder}}
 (Clarendon Press, Oxford, 1982).

\bibitem{FTT-1982}
 E.~M.~Ganapolskii and D.~N.~Makovetskii,
 Sov. Phys. - Solid State. {\textbf{24}}, 1119 (1982)
 [translation of: Fiz. Tverd. Tela (Leningrad) {\textbf{24}},
 1960 (1982)].


\end{thebibliography}
\end{document}